# Magnetic-dipolar and electromagnetic vortices in quasi-2D ferrite disks

M. Sigalov, E.O. Kamenetskii, and R. Shavit

Ben-Gurion University of the Negev, Beer Sheva 84105, Israel

June 19, 2008

**Abstract**

Magnetic-dipolar-mode (MDM) oscillations in a quasi-2D ferrite disk show unique dynamical symmetry properties resulting in appearance of topologically distinct structures. Based on the magnetostatic (MS) spectral problem solutions, in this paper we give an evidence for eigen MS power-flow-density vortices in a ferrite disk. Due to these circular eigen power flows, the MDMs are characterized by MS energy eigen states. It becomes evident that the reason of stability of the vortex configurations in saturated ferrite samples is completely different from the nature of stability in magnetically soft cylindrical dots. We found a clear correspondence between analytically derived MDM vortex states and numerically modeled electromagnetic vortices in quasi-2D ferrite disks.

## I. INTRODUCTION

In ferromagnetic systems, one can clearly distinguish three characteristic scales. There are the scales of the spin (exchange interaction) fields, the magnetostatic (dipole-dipole interaction) fields, and the electromagnetic fields. These fields may define different vortex states, which are characterized by different kinds of physical phenomena and are observed in ferromagnetic samples with different sizes. Together with magnetization vortices in magnetically soft samples [1 – 3] and the magnetostatic (MS) vortex behaviors in saturated ferrite disks [4], one can observe electromagnetic vortices originated from ferrite samples in microwave cavities [5 – 8]. In the last case, vortices appear because of the time-reversal-symmetry-breaking (TRSB) effects resulting in a rich variety of the electromagnetic wave topological phenomena [5 – 8]. In spite of the fact that vortices are observed in different kinds of physical phenomena, yet such "swirling" entities seem to elude an all-inclusive definition. In quite a number of problems one can define a vortex as a circular flow which is attributed with a certain phase factor and a circular integral of a gradient of the phase gives a non-zero quantity. This quantity is multiple to a number of full rotations.

In a magnetically soft cylindrical dot of micrometer or submicrometer size, the vortex structures are stable because of competition between the exchange and dipole interactions [1 – 3]. These stable vortex configurations are considered as very attractive objects for fundamental physics studies and for potential use as a unit cell for high density magnetic storage and magnetic logic devices. The vortex structures in magnetically saturated cylindrical samples are characterized by negligibly small exchange fluctuations [4]. It becomes evident that the reason of stability of the vortex configurations in saturated dots should be completely different from the nature of stability in magnetically soft cylindrical dots. The magnetic dipole interaction provides us with a long-range mechanism of interaction, where a magnetic medium is considered as a continuum. It is well known that magnetostatic (MS) ferromagnetism has a character essentially different from exchange ferromagnetism [9, 10]. This statement finds a strong confirmation in confinement phenomena of magnetic-dipolar-mode (MDM) (or MS) oscillations. MDM problems border with the exchange-interaction magnetization dynamics, on one hand, and with the electromagnetic behaviors, on the other hand. The feature of MDM oscillations in small samples with strong temporal dispersion of the permeability tensor: $\ddot{\mu} = \ddot{\mu}(\omega)$, is the fact that one neglects variation of the exchange energy and variations of the electric energy [11]. In the attempts to consider magnetic-dipolar interactions in ferrite samples with the dipolar and exchange energy competition, one usually solves "pure static" magnetostatic (MS) equations for the dipolar field [12 – 14]. It should be taken into account, however, that the magnetic-dipolar interaction provides us with a long-range mechanism of interaction, where a magnetic medium is considered as a continuum and to get correct solutions for MS dynamical processes inside a ferrite particle one has to presuppose existence of certain *retardation effects* for the MS fields. The magnetic samples which exhibit the MS resonance behaviors in microwaves are described by the MS-potential *wave functions* $\psi(\vec{r},t)$.

A recently published spectral theory of MDMs in quasi-2D ferrite disks [4, 15, 16] gives a deep insight into an explanation of the experimental MS multiresonance absorption spectra shown both in well known previous [17, 18] and new [19, 20] studies. The theory suggests an existence of othogonality relations for the MS-potential eigen modes. In a quasi-2D ferrite disk, a differential equation for this scalar complex wave function resembles the Schrödinger equation. It follows that the MDM spectral problem in a normally magnetized ferrite disk is characterized by energy eigenstates [15, 16]. The energy orthogonality for MDMs in a ferrite disk is accompanied with the dynamical symmetry breaking effects resulting in appearance of the vortex structures [4]. This shows that the vortex states can be created not only in magnetically soft "small" (with the dipolar and exchange energy competition) cylindrical dots, but also in magnetically saturated "big" (when the exchange fluctuations can be neglected) cylindrical samples.

Together with the exchange-interaction and magnetic-dipolar vortex behaviors, electromagnetic (Poynting-vector) vortices can be observed in ferromagnetic systems. A ferrite is a magnetic dielectric with low losses. This may allow for electromagnetic waves to penetrate the ferrite and results in an effective interaction between the electromagnetic waves and magnetization within the ferrite. At the ferromagnetic resonance conditions (FMR), such an interaction can demonstrate very interesting physical features. Because of inserting a piece of a magnetized ferrite into the resonator domain, a microwave resonator behaves under odd-time-reversal symmetry. Solutions of the fields in microwave systems with enclosed ferrite samples are considered as solutions of non-integrable Maxwellian problems. In this case a ferrite sample may act as a topological defect causing induced electromagnetic vortices. It was shown numerically [6] that the power-flow lines of the microwave-cavity field interacting with a ferrite sample, in the proximity of its FMR, forms whirlpool-like electromagnetic vortices. It was found that in such non-integrable structures, magnetic gyrotropy and geometrical factors lead to special effects of dynamical symmetry breaking resulting in effective chiral magnetic currents and topological magnetic charges [7, 8].

Based on the magnetostatic (MS) spectral problem solutions, in this paper we show the presence of eigen MS power-flow-density vortices in a ferrite disk. Such circular eigen power flows give evidence for the MDM energy eigen states. This results in stability of the vortex configurations in magnetically saturated dots. We analyze the MDM fields in a ferrite sample. Based on a numerical analysis we show that for certain geometry of a ferrite disk and certain FMR frequency and bias-magnetic-field regions one has strong pronounced *eigenfunction patterns* of the topologically distinct electromagnetic vortex structures. The spectral properties of MDM oscillations in a ferrite disk cannot be analytically described based on the complete-set Maxwell equations. So the dynamical processes for the MS fields arising from the MS-potential-wave-function description cannot be formally considered as (and reduced to) the effects obtained from the complete-set Maxwell equations. The Maxwell-equation mode eigenstates in a ferrite disk are numerically modeled objects in the ray phase space. The fact that such non-integrable-electromagnetic-problem solutions are in clear correspondence with the analytical spectral solutions for the MDM vortex states in quasi-2D ferrite disks is one of the most interesting phenomena shown in this paper.

## II. MDM VORTICES IN QUASI-2D FERRITE DISKS

### A. MDM spectral problem in a quasi-2D ferrite disk

Generally, in classical electromagnetic problem solutions for time-varying fields, there are no differences between the methods of solutions: based on the electric- and magnetic-field representation or based on the potential representation of the Maxwell equations. For the wave processes, in the field representation we solve a system of first-order partial differential equations (for the electric and magnetic vector fields) while in the potential representation we have a smaller number of second-order differential equations (for the scalar electric or vector magnetic potentials). The potentials are introduced as formal quantities for a more convenient way to solve the problem and a set of equations for potentials are equivalent in all respects to the Maxwell equations for fields [21]. The situation can become completely different if one supposes to solve the electromagnetic boundary problem for wave processes in small samples of a strongly temporally dispersive magnetic medium [11]. For such magnetic samples with the MS resonance behaviors in microwaves, the problem cannot be formally reduced to the complete-set Maxwell-equation representation [4, 15, 16] and one becomes faced with a special role of the MS-potential wave function $\psi(\vec{r},t)$. This scalar wave function acquires a physical meaning in the MDM spectral problem and results in experimental observation of energy shifts of oscillating modes and eigen electric moments [17 – 20].

For a quasi-2D ferrite disk with cylindrical coordinates $z, r, \theta$, one uses separation of variables for MS-potential wave function and a spectral problem is formulated with respect to membrane MS functions (described by coordinates $r, \theta$). In solving the spectral problem for MDM oscillations, one deals with two types of differential-operator equations. The first type is

$$\left( \hat{L}_\perp - i\beta \, \hat{R} \right) \begin{pmatrix} \vec{\tilde{B}} \\ \tilde{\varphi} \end{pmatrix} = 0, \qquad (1)$$

where $\tilde{\varphi}$ is a dimensionless membrane MS-potential wave function, $\vec{\tilde{B}}$ is a dimensionless membrane function of a magnetic flux density. For a normally magnetized disk characterizing by a diagonal, $\mu$, and off-diagonal, $\mu_a$, components of the permeability tensor $\vec{\mu}$, the components of vector $\vec{\tilde{B}}$ are: $\tilde{B}_r = \mu \frac{\partial \tilde{\varphi}}{\partial r} + i\mu_a \frac{\partial \tilde{\varphi}}{\partial \theta}$ and $\tilde{B}_\theta = \mu \frac{\partial \tilde{\varphi}}{\partial \theta} - i\mu_a \frac{\partial \tilde{\varphi}}{\partial r}$. In Eq. (1), $\hat{L}_\perp$ is a differential-matrix operator:

$$\hat{L}_\perp \equiv \begin{pmatrix} (\vec{\mu}_\perp)^{-1} & \nabla_\perp \\ -\nabla_\perp \cdot & 0 \end{pmatrix}, \qquad (2)$$

(subscript $\perp$ means correspondence with the in-plane, $r, \theta$, coordinates), $\beta$ is the MS-wave propagation constant along $z$ axis ($\varphi \equiv \tilde{\varphi} \, e^{-i\beta z}$), and $\hat{R}$ is a matrix:

$$\hat{R} \equiv \begin{pmatrix} 0 & \vec{e}_z \\ -\vec{e}_z & 0 \end{pmatrix}, \qquad (3)$$

where $\vec{e}_z$ is a unit vector along $z$ axis. Operator $\hat{L}_\perp$ becomes self-adjoint for homogeneous boundary conditions (continuity of $\tilde{\varphi}$ and $\tilde{B}_r$) on a lateral surface of a ferrite disk.

The second type of a differential-operator equation is

$$\left(\hat{G}_\perp - (\beta)^2\right)\tilde{\eta} = 0, \tag{4}$$

where $\tilde{\eta}$ is a dimensionless membrane MS-potential wave function (different from function $\tilde{\varphi}$),

$$\hat{G}_\perp \equiv \mu \nabla_\perp^2, \tag{5}$$

and $\nabla_\perp^2$ is the two-dimensional (with respect to in-plane coordinates) Laplace operator. Operator $\hat{G}_\perp$ is positive definite for negative quantities $\mu$. Outside a ferrite region Eq. (4) becomes the Laplace equation ($\mu = 1$). Double integration by parts on square $S$ – an in-plane cross-section of a disk structure – of the integral $\int_S (\hat{G}_\perp \tilde{\eta})\tilde{\eta}^* dS$ gives the boundary conditions for self-adjointess of operator $\hat{G}_\perp$.

MS-potential wave functions $\tilde{\eta}$ and $\tilde{\varphi}$ are different one from another. In a cylindrical coordinate system we can write $\tilde{\eta} = \tilde{\eta}(r)\tilde{\eta}(\theta)$ and $\tilde{\varphi} = \tilde{\varphi}(r)\tilde{\varphi}(\theta)$. For angular parts we have $\tilde{\eta}(\theta) \sim e^{-i\nu\theta}$ and $\tilde{\varphi}(\theta) \sim e^{-i\nu\theta}$, where $\nu = \pm 1, \pm 2, \pm 3 \ldots$. For self-adjointess of operator $\hat{G}_\perp$ one has the homogeneous boundary condition for a disk of radius $\Re$:

$$\mu\left(\frac{\partial \tilde{\eta}}{\partial r}\right)_{r=\Re^-} - \left(\frac{\partial \tilde{\eta}}{\partial r}\right)_{r=\Re^+} = 0. \tag{6}$$

At the same time, for self-adjointess of operator $\hat{L}_\perp$ the homogeneous boundary condition is:

$$\mu\left(\frac{\partial \tilde{\varphi}}{\partial r}\right)_{r=\Re^-} - \left(\frac{\partial \tilde{\varphi}}{\partial r}\right)_{r=\Re^+} = -\frac{\mu_a}{\Re}\nu(\tilde{\varphi})_{r=\Re^-}. \tag{7}$$

Functions $\tilde{\eta}(r)$ and $\tilde{\varphi}(r)$ are the Bessel functions and the boundary conditions (6) and (7) can be represented, respectively, as

$$(-\mu)^{\frac{1}{2}}\frac{J_\nu'}{J_\nu} + \frac{K_\nu'}{K_\nu} = 0 \tag{8}$$

and

$$(-\mu)^{\frac{1}{2}}\frac{J_\nu'}{J_\nu} + \frac{K_\nu'}{K_\nu} - \frac{\mu_a \nu}{|\beta|\Re} = 0. \tag{9}$$

Here $J_\nu$, $J'_\nu$, $K_\nu$, and $K'_\nu$ are the values of the Bessel functions of order $\nu$ and their derivatives (with respect to the argument) on a lateral cylindrical surface ($r = \Re$). From boundary condition (7) [or (9)] it evidently follows that for an integer azimuth number $\nu$ and a given sign of parameter $\mu_a$, there are different functions, $\tilde{\varphi}_+$ and $\tilde{\varphi}_-$, for positive and negative directions of an angle coordinate when $0 \leq \theta \leq 2\pi$. So function $\tilde{\varphi}$ is not a single-valued function. It changes a sign when $\theta$ is rotated by $2\pi$. At the same time, function $\tilde{\eta}$ is a single-valued function.

For a given magnetic-dipolar mode, function $\tilde{\varphi}$ is normalized on a power flow density along $z$ axis. It is well known that in an axially magnetized infinite ferrite rod, there exist two magnetostatic waves propagating along a positive direction of $z$ axis and corresponding to the $\tilde{\varphi}_+$ and $\tilde{\varphi}_-$ solutions [22]. These two waves are degenerate with respect to an average density of an accumulated energy, which is a norm for function $\tilde{\eta}$. This fact has a formal aspect for MS-wave propagation in an infinite ferrite rod, but acquires, however, a peculiar physical meaning in a ferrite disk [4, 15, 16]. In a spectral problem of MDMs in a normally magnetized quasi-2D ferrite disk, Eq. (6) [or Eq. (8)] describe so-called *essential boundary conditions* while Eq. (7) [or Eq. (9)] describe so-called *natural boundary conditions* [16, 23]. A specific boundary relation on a lateral surface of a ferrite disk gives a clear correspondence between double-valued functions $\tilde{\varphi}$ and single-valued functions $\tilde{\eta}$ [4]:

$$\left(\tilde{\varphi}_\pm\right)_{r=\Re} = \delta_\pm \left(\tilde{\eta}\right)_{r=\Re}, \tag{10}$$

where

$$\delta_\pm \equiv f_\pm e^{-iq_\pm \theta} \tag{11}$$

is a double-valued edge (at $r = \Re$) function. The azimuth number $q_\pm$ is equal to $\pm \frac{1}{2}$ and for amplitudes we have $f_+ = -f_-$ and $|f_\pm| = 1$. Because of the boundary relation (10), the topological effects are manifested by the generation of relative phases which accumulate on the edge wave function $\delta_\pm$. These topological effects become apparent through the integral fluxes of the pseudo-electric fields [4]. There should be the positive and negative fluxes corresponding to counterclockwise and clockwise edge-function chiral rotation. The different-sign fluxes are inequivalent to avoid cancellation. Every MDM in a thin ferrite disk is characterized by a certain energy eigenstate and two different-sign fluxes of the pseudo-electric fields which are energetically degenerate.

**B. Power-flow-density vortices and energy eigenstates in a quasi-2D ferrite disk with MDM oscillations**

For MS wave propagating in a ferrite rod, membrane function $\tilde{\varphi}$ is normalized on a power flow density along $z$ axis. This is an observable quantity. For MDM oscillations in a ferrite disk, power flow density along $z$ axis is evidently equal to zero. One of the unique features of such oscillations is the presence of azimuth power flow densities.

In a general representation, for monochromatic MS-wave processes with time variation $\sim e^{i\omega t}$, the power flow density (in Gaussian units) is expressed as [4]:

$$\vec{p} = \frac{i\omega}{16\pi}\left(\psi^* \vec{B} - \psi \vec{B}^*\right). \tag{12}$$

The proof that this is really a power flow density arises from the following equation:

$$-\frac{i\omega}{16\pi}\nabla\cdot\left(\psi\vec{B}^*-\psi^*\vec{B}\right)=-\frac{i\omega}{16\pi}\left[\vec{B}\cdot\left(\vec{\mu}^*(\omega)\right)^{-1}\cdot\vec{B}^*-\vec{B}^*\cdot\left(\vec{\mu}(\omega)\right)^{-1}\cdot\vec{B}\right]. \tag{13}$$

This is, in fact, the energy balance equation for monochromatic MS waves in lossy magnetic media. In the right-hand side of this equation we have the average density of magnetic losses taken with an opposite sign. Thus the term in the left-hand side is the divergence of the power flow density.

The power flow density for a certain magnetic-dipolar mode $n$ is

$$\vec{p}_n=\frac{i\omega}{16\pi}\left(\psi_n^*\vec{B}_n-\psi_n\vec{B}_n^*\right). \tag{14}$$

For MS-potential wave function for mode $n$ one has [4, 15, 16]

$$\psi_n=C_n\xi_n(z)\tilde{\varphi}_n(r,\theta), \tag{15}$$

where $\xi_n(z)$ is an amplitude factor, $C_n$ is a dimensional coefficient, and $\tilde{\varphi}_n(r,\theta)$ is a dimensionless membrane function. For magnetic flux density ($\vec{B}_n=-\vec{\mu}\cdot\nabla\psi_n$), we can write

$$\vec{B}_n=(B_n)_z\vec{e}_z+\tilde{B}_n\vec{e}_\perp, \tag{15}$$

where

$$(B_n)_z=-C_n\frac{\partial\xi_n(z)}{\partial z}\tilde{\varphi}_n(r,\theta) \tag{16}$$

and

$$\tilde{B}_n=-C_n\xi_n(z)\left[\vec{\mu}_\perp\cdot\vec{\nabla}_\perp\tilde{\varphi}_n(r,\theta)\right]\cdot\vec{e}_\perp. \tag{17}$$

Subscript $\perp$ corresponds to transversal (with respect to $z$ axis) components.

For oscillating MDMs in a quasi-2D ferrite disk not only the $z$ component of the power flow density is equal to zero. It is easy to show that the $r$ component of the power flow density is equal to zero as well. There is, however, non-zero real azimuth component of the power flow density:

$$\left(p_n(z)\right)_\theta=\frac{i\omega}{16\pi}C_n^2(\xi(z))^2\left[-\mu\frac{1}{r}\left(\tilde{\varphi}_n^*\frac{\partial\tilde{\varphi}_n}{\partial\theta}-\tilde{\varphi}_n\frac{\partial\tilde{\varphi}_n^*}{\partial\theta}\right)+i\mu_a\left(\tilde{\varphi}_n^*\frac{\partial\tilde{\varphi}_n}{\partial r}+\tilde{\varphi}_n\frac{\partial\tilde{\varphi}_n^*}{\partial r}\right)\right]. \tag{18}$$

With use of representation $\tilde{\varphi}_n=\tilde{\varphi}_n(r)\tilde{\varphi}_n(\theta)$, where $\tilde{\varphi}_n(\theta)\sim e^{-i\nu_n\theta}$ and $\nu_n$ is an integer, one has

$$\left(p_n(r,z)\right)_\theta=\frac{\tilde{\varphi}_n(r)}{8\pi}\omega C_n^2(\xi_n(z))^2\left[-\tilde{\varphi}_n(r)\frac{\mu}{r}(\nu_n)-\mu_a\frac{\partial\tilde{\varphi}_n(r)}{\partial r}\right]. \tag{19}$$

This is a non-zero circulation quantity around a circle $2\pi r$. An amplitude of a MS-potential function is equal to zero at $r=0$. For a scalar wave function, this presumes the Nye and Berry phase singularity [24]. Circulating quantities $(p_n(r,z))_\theta$ are the MDM power-flow-density vortices with cores at the disk center. At a vortex center amplitude of $(p_n)_\theta$ is equal to zero. It follows from Eq. (19) that for a given mode number $n$ characterizing by a certain function $\tilde{\varphi}_n(r)$ there will be different power flow densities $(p_n(r,z))_\theta$ for different signs of the azimuth number $\nu_n$.

For calculation of the power flow density, the wave number $\beta$ and functions $\xi(z)$ and $\tilde{\varphi}(r)$ for a given mode can be found based on solution of a system of two equations: a characteristic equation for MS waves in an axially magnetized ferrite rod [Eq. (9)] and a characteristic equation for MS waves in a normally magnetized ferrite slab:

$$\tan(\beta h) = -\frac{2\sqrt{-\mu}}{1+\mu}, \qquad (20)$$

where $h$ is a disk thickness [15, 16].

Stable vortex structures of the power flow density (PFD) for every oscillating MDM give an evidence for energy eigenstates in a quasi-2D ferrite disk. Since in a lossless ferrite the divergence of the PFD vortex is equal to zero, the only possibility to change the PFD topological structure is via discrete energy transition between the MDM energy levels. The statement that confinement phenomena for MS oscillations in a normally magnetized ferrite disk demonstrate typical atomic-like properties of discrete energy levels becomes evident from an analysis of experimental absorption spectra. The main feature of multi-resonance line spectra observed in well known experiments [17, 18] is the fact that high-order peaks correspond to lower quantities of the bias DC magnetic field. Physically, the situation looks as follows. Let $H_0^{(A)}$ and $H_0^{(B)}$ be, respectively, the upper and lower values of a bias magnetic field corresponding to the borders of a region. We can estimate a total depth of a "potential well" as: $\Delta U = 4\pi M_0 \left( H_0^{(A)} - H_0^{(B)} \right)$, where $M_0$ is the saturation magnetization. Let $H_0^{(1)}$ be a bias magnetic field, corresponding to the main absorption peak in the experimental spectrum ($H_0^{(B)} < H_0^{(1)} < H_0^{(A)}$). When we put a ferrite sample into this field, we supply it with the energy: $4\pi M_0 H_0^{(1)}$. To some extent, this is a pumping-up energy. Starting from this level, we can excite the entire spectrum from the main mode to the high-order modes. As a value of a bias magnetic field decreases, the "particle" obtains the higher levels of negative energy. One can estimate the negative energies necessary for transitions from the main level to upper levels. For example, to have a transition from the first level $H_0^{(1)}$ to the second level $H_0^{(2)}$ ($H_0^{(B)} < H_0^{(2)} < H_0^{(1)} < H_0^{(A)}$) we need the density energy surplus: $\Delta U_{12} = 4\pi M_0 \left( H_0^{(1)} - H_0^{(2)} \right)$. The situation is very resembling the increasing a negative energy of the hole in semiconductors when it "moves" from the top of a valence band. In a classical theory, negative-energy solutions are rejected because they cannot be reached by a continuous loss of energy. But in quantum theory, a system can jump from one energy level to a discretely lower one; so the negative-energy solutions cannot be rejected, out of hand. When one continuously varies the quantity of the DC field $H_0$, for a given quantity of frequency $\omega$, one sees a discrete set of absorption peaks. It means that one has the discrete-set levels of potential energy. The line spectra appear due to the quantum-like transitions between energy levels of a ferrite disk-form particle.

The statement of stable vortex structures of the PFD and energy eigenstates of MDMs reveals some discrepancies for MS-potential functions on a lateral surface of a ferrite disk. To settle the problem, one should impose a certain phase factor [4]. The solutions for MS-potential functions $\tilde{\varphi}$ with additional phase factors satisfy the mode orthogonality relations. Such boundary phase factors cause appearance of effective surface circular magnetic currents on a lateral surface of a disk [4]. It appears that circular eigen power flows are accompanied with the fluxes of the pseudo-electric fields. These fluxes are induced so that the edge wave function $\delta$ provides necessary boundary conditions to achieve singlevaluedness of function $\tilde{\eta}$ [see Eq. (10)]. This guaranties energy eigenstates of MDMs in a ferrite disk.

All the energy relations have physical meaning when are written for a single-valued membrane function $\tilde{\eta}(r,\theta)$. For MDM oscillations one has energy eigenstates which are characterized by a two-dimensional ("in-plane") differential operator [4, 15, 16]

$$\hat{F}_\perp = \frac{g_q}{16\pi} \mu \nabla_\perp^2, \qquad (21)$$

where $g_q$ is a dimensional normalization coefficient for mode $q$. The normalized average (on the RF period) density of accumulated magnetic energy of mode $q$ is determined as

$$E_n = \frac{g_q}{16\pi} (\beta_n)^2. \qquad (22)$$

The energy eigenvalue problem is defined by the differential equation:

$$\hat{F}_\perp \tilde{\eta}_n = E_n \tilde{\eta}_n, \qquad (23)$$

For MDMs in a ferrite disk at a constant frequency one has the energy orthonormality:

$$(E_n - E_{n'}) \int_S \tilde{\eta}_n \tilde{\eta}_{n'}^* dS = 0, \qquad (24)$$

where $S$ is a cylindrical cross section of an open disk. Different mode energies one has at different quantities of a bias magnetic field. From the principle of superposition of states, it follows that wave functions $\tilde{\eta}_n$ ($n = 1, 2, ...$), describing our "quantum" system, are "vectors" in an abstract space of an infinite number of dimensions – the Hilbert space. The scalar-wave membrane function $\tilde{\eta}$ can be represented as

$$\tilde{\eta} = \sum_n a_n \tilde{\eta}_n \qquad (25)$$

and the probability to find a system in a certain state $q$ is defined as

$$|a_n|^2 = \left| \int_S \tilde{\eta} \tilde{\eta}_n^* dS \right|^2. \qquad (26)$$

It is very important to note that for the energy eigenvalue problem, wavenumber $\beta_n$ in Eq. (22) and the mode structure for $\tilde{\eta}_n$ are defined based on solutions of a system of two equations: Eq. (8) and Eq. (20).

**C. The field structures in a quasi-2D ferrite disk with MDM vortices**

The MDM vortices in a quasi-2D ferrite disk are characterized by specific field structures. For non-uniform ferromagnetic resonances in small ferrite samples there are no electromagnetic retardation effects since one neglects electric displacement currents [11]. It means that the electric and magnetic fields in our case of a quasi-2D ferrite disk cannot be obtained analytically from the Maxwell-equation *spectral* problem. At the same time, the fields can be derived from the MDM *spectral* problem solved for the MS-potential wave functions $\psi(\vec{r},t)$. So in the ferrite-disk boundary value problem, the MS-potential wave function is a primary notion while the electric and magnetic fields can be related to a secondary concept.

Inside a ferrite disk ($r \leq \Re$, $0 \leq z \leq h$) one has:

$$\psi(r,\theta,z) = J_\nu\left(\frac{\beta r}{\sqrt{-\mu}}\right)\left(\cos\beta z + \frac{1}{\sqrt{-\mu}}\sin\beta z\right)e^{-j\nu\theta}. \tag{27}$$

This function satisfies characteristic equations (9) and (20). Based on such a MS-potential function one defines the magnetic field ($\vec{H} = -\vec{\nabla}\psi$) inside a ferrite disk as

$$H_r(r,\theta,z) = \frac{\beta}{\sqrt{-\mu}} J'_\nu\left(\frac{\beta r}{\sqrt{-\mu}}\right)\left(\cos\beta z + \frac{1}{\sqrt{-\mu}}\sin\beta z\right)e^{-i\nu\theta}, \tag{28}$$

$$H_\theta(r,\theta,z) = \frac{-i\theta}{r} J_\nu\left(\frac{\beta r}{\sqrt{-\mu}}\right)\left(\cos\beta z + \frac{1}{\sqrt{-\mu}}\sin\beta z\right)e^{-i\nu\theta}, \tag{29}$$

$$H_z(r,\theta,z) = \beta J_\nu\left(\frac{\beta r}{\sqrt{-\mu}}\right)\left(-\sin\beta z + \frac{1}{\sqrt{-\mu}}\cos\beta z\right)e^{-i\nu\theta}. \tag{30}$$

In a small sample of a material with a strong temporal dispersion of permeability, differential equations for the magnetic field and for the magnetic flux density are the same as the corresponding equations for "pure" magnetostatics. The only (and substantial) difference is that the permeability is not a constant scalar quantity but is a tensor with the components strongly dependent on frequency. In this case, a role of the electric field in the wave process is not clearly determined. Formally, one can suppose that for the monochromatic MS-wave process there exists a curl electric field $\vec{E}$ defined by the Faraday law. One can represent the electric field as follows:

$$\vec{\nabla}\times\vec{E} = -\frac{i}{c}\omega\vec{B} = -\frac{i}{c}\omega\vec{H} - \frac{i}{c}4\pi\omega\vec{m}, \tag{31}$$

where $\vec{m}$ is RF magnetization. With the $\vec{\nabla}\times$ differential operation for the left-hand and right-hand sides of Eq. (31) and taking into account that $\vec{\nabla}\times\vec{H} = 0$, one obtains:

$$\nabla^2\vec{E} = \frac{i}{c}4\pi\omega\nabla\times\vec{m}. \tag{32}$$

Here we used the relation $\nabla \cdot \vec{E} = 0$. This follows from the fact that in the MS-wave description no electric polarization effects are taken into consideration.

The electric field can be formally represented as being originated from an effective electric current:

$$\nabla^2 \vec{E} = i\omega \frac{4\pi}{c^2} \vec{j}^{(e)}, \tag{33}$$

where

$$\vec{j}^{(e)} \equiv c\vec{\nabla} \times \vec{m}. \tag{34}$$

Taking into account that $\vec{m} = \vec{\tilde{\chi}} \cdot \vec{H}$, where the magnetic susceptibility [26]

$$\vec{\tilde{\chi}} = \begin{bmatrix} \chi & i\chi_a & 0 \\ -i\chi_a & \chi & 0 \\ 0 & 0 & 0 \end{bmatrix}, \tag{35}$$

we have the following components of an effective electric current:

$$j_r^{(e)} = ic\left(\frac{\chi_a \beta^2}{\sqrt{-\mu}} J_\nu'\left(\frac{\beta r}{\sqrt{-\mu}}\right) + \frac{\beta \chi \nu}{r} J_\nu\left(\frac{\beta r}{\sqrt{-\mu}}\right)\right)\left(\sin \beta z - \frac{1}{\sqrt{-\mu}} \cos \beta z\right)e^{-i\nu\theta}, \tag{36}$$

$$j_\theta^{(e)} = c\left(\frac{\chi \beta^2}{\sqrt{-\mu}} J_\nu'\left(\frac{\beta r}{\sqrt{-\mu}}\right) + \frac{\beta \chi_a \nu}{r} J_\nu\left(\frac{\beta r}{\sqrt{-\mu}}\right)\right)\left(\sin \beta z - \frac{1}{\sqrt{-\mu}} \cos \beta z\right)e^{-i\nu\theta}, \tag{37}$$

$$j_z^{(e)} = -ic \frac{1}{4\pi} \frac{\mu_a}{\mu} J_\nu\left(\frac{\beta r}{\sqrt{-\mu}}\right)\left(\cos \beta z + \frac{1}{\sqrt{-\mu}} \sin \beta z\right)e^{-i\nu\theta}. \tag{38}$$

In the MDM vortex structure, a magnetic field described by Eqs. (28) – (30) and an effective electric current described by Eqs. (36) – (37) [and therefore an electric field described by Eq. (33)] show the running azimuth wave behaviors. By multiplying these equations at $e^{i\omega t}$ and taking real parts, one has the real-time azimuth waves.

The electric field determined from the above consideration has an auxiliary character for MS-wave processes. For a curl electric field one can introduce a magnetic vector potential: $\vec{E} \equiv -\nabla \times \vec{A}^m$. Based on the Faraday law and taking into account that $\vec{B} = -\vec{\mu}(\omega) \cdot \nabla \psi$, we have

$$\nabla^2 \vec{A}^m - \nabla(\nabla \cdot \vec{A}^m) - \frac{i\omega}{c} \vec{\mu}(\omega) \cdot \nabla \psi = 0. \tag{39}$$

This equation shows that formally two types of gauges are possible. In the first type of a gauge we have:

$$\nabla \cdot \vec{A}^m = 0 \tag{40}$$

and, therefore,

$$\nabla^2 \vec{A}^m = \frac{i\omega}{c} \vec{\mu}(\omega) \cdot \nabla \psi \,. \tag{41}$$

The second type of a gauge is written as

$$\nabla(\nabla \cdot \vec{A}^m) + \frac{i\omega}{c} \vec{\mu}(\omega) \cdot \nabla \psi = 0 \tag{42}$$

and, therefore,

$$\nabla^2 \vec{A}^m = 0 \,. \tag{43}$$

The last equation shows that any sources of the electric field are not defined and thus the electric field is not defined at all. So only the first type of a gauge, giving Eq. (41), should be taken into account. The main point, however, is that the considered above gauge transformation does not fall under the known gauge transformations, neither the Lorentz gauge nor the Coulomb gauge [21], and cannot formally lead to the wave equation. Moreover, to have a wave process one should assume that there exists a certain non-physical mechanism describing the effect of transformation of the curl ($\vec{E} = -\nabla \times \vec{A}^m$) electric field to the potential ($\vec{H} = -\nabla \psi$) magnetic field.

Analytically described MDM spectral properties are well verified in microwave experiments. When a thin-film ferrite disk is placed in a microwave cavity, one observes experimentally the energy shifts and eigen electric moments of oscillating MDMs [17 – 20]. An analysis of excitation of the MDM vortices in a ferrite disk by external electromagnetic fields is, however, beyond the frames of analytical solutions. Electromagnetically, a microwave cavity with an inserted ferrite disk is a non-integrable system. It appears that for a very thin ferrite disk and for a certain region of external parameters (a working frequency and a bias magnetic field), numerical simulation shows a set of electromagnetic modes inside a disk. These modes are characterized by the PFD vortex states. It is very surprisingly to find that the structures of the PFD vortices and the field patterns for these numerically modeled electromagnetic modes are in excellent correspondence with the analytically derived PFD vortices and the field patterns of MDMs in a ferrite disk.

## III. ELECTROMAGNETIC VORTICES AND TOPOLOGICAL RESONANT STATES IN CAVITIES WITH THIN-FILM FERRITE DISKS

In solving Maxwell's equations, one of the powerful approaches is the short-wavelength approximation leading to the ray picture. Rays are the solutions of the Fermat's variational principle, which in particular implies the laws of reflection and refraction at interfaces. As soon as one makes the transition from wave physics to the classical ray dynamics, concepts such as "trajectory" and "phase space" become meaningful. For microwave and optical cavities one may use a model of the geometrical rays within the boundaries – the billiard model. By changing shape parameters in billiard models, it is possible to describe systems with classical motion ranging from integrable to fully chaotic. For example, the dynamics of a classical particle (a classical ray) bouncing on a plane between hard walls depends in a characteristic way on a shape of the billiard. Its motion in a rectangular or circular billiard is regular and the system is integrable. However, if a circular obstacle is put inside the rectangle the system becomes chaotic. In a case of a metal circular obstacle one has the Sinai billiard. On the other hand, if an integrable system of a rectangular or circular billiard is transformed into a stadium, one has a chaotic system of the Bunimovich billiard.

Integrable systems are characterized by a complete set of "good quantum numbers". If, however, the ray system is not integrable then the quantization procedure fails. It is not known in this case what the correspondence could be between normal modes of the wave system and objects in the ray phase space, if indeed there is any correspondence at all. The distinction between integrable and non-integrable classical dynamical systems has the qualitative implication of regular motion vs. chaotic motion. A chaotic system cannot be decomposed; the motion along one coordinate axis is always coupled to what happens along the other axis. In a case of classical non-integrable problems, the corresponding wave equation cannot be reduced to a set of mutually independent ordinary differential equations. Their coupling makes it impossible to label the wave solutions by a complete set of eigen numbers. At the same time, because of a clear correspondence between the two-dimensional Helmholtz equation and the Schrödinger equation, an analysis of non-integrable planar microwave and optical cavities can be made based on the theories of quantum chaos. A superposition of random plane waves can be used to describe the statistics of chaotic wave functions (the so-called "wave chaos") in electromagnetic cavities [25].

Microwave cavities with ferrite inclusions can be modeled as billiards with the time reversal symmetry breaking effects. Because of the time-reversal symmetry breaking effects, a system of a rectangular-waveguide cavity with an embedded inside ferrite disk (even having sizes much small compared with the cavity sizes) is not a weakly perturbed integrable system, but a non-integrable system [5 – 8]. At the same time, electromagnetic wave processes inside a ferrite sample for a constant bias field can be approximated as chaotically propagating plane waves. When a bias magnetic field is directed along $z$ axis, a ferrite material is described by the permeability tensor

$$\vec{\vec{\mu}} = \begin{bmatrix} \mu & j\mu_a & 0 \\ -j\mu_a & \mu & 0 \\ 0 & 0 & 1 \end{bmatrix}, \tag{44}$$

and the plane-wave propagation in an infinite ferrite medium is characterized by the following scalar effective permeability parameter [26]:

$$\mu_{\mathit{eff}}(\theta_k) = \frac{2+\sin^2\theta_k(\mu_\perp - 1) \pm \sqrt{\sin^4\theta_k(\mu_\perp - 1)^2 + 4\cos^2\theta_k\,\mu_a^2/\mu^2}}{2(\sin^2\theta_k + \cos^2\theta_k/\mu)}, \tag{45}$$

where $\theta_k$ is an angle between directions of the wave vector $\vec{k}$ and the bias magnetic field $\vec{H}_0$, and

$$\mu_\perp \equiv \frac{\mu^2 - \mu_a^2}{\mu}. \tag{46}$$

Two limit cases have to be taken into account: for $\theta_k = \frac{\pi}{2}$ one has an effective permeability parameter $\mu_{\mathit{eff}} = \mu_\perp$ and for $\theta_k = 0$ there are two effective permeability parameters

$$\mu_{\mathit{eff}_+} = \mu + \mu_a \quad \text{and} \quad \mu_{\mathit{eff}_-} = \mu - \mu_a, \tag{47}$$

characterizing, respectively, the plane waves with right-hand and left-hand circular polarizations. Only for the right-hand circularly polarized wave, one can distinguish two behaviors: $\mu_{\mathit{eff}_+} > 0$ and $\mu_{eff_+} < 0$. The quantity $\mu_{eff_-}$ is always positive [26]. Quantities and signs of components of the permeability tensor (and therefore quantities and signs of effective permeabilities $\mu_\perp$ and $\mu_{eff_\pm}$) are dependent on a frequency, a bias magnetic field and on a saturation magnetization parameter of a ferrite.

Let us consider separately plane electromagnetic waves propagating perpendicular to a bias magnetic field and parallel to a bias magnetic field in a lossless ferrite medium. In our studies of the electromagnetic fields inside a normally magnetized ferrite disk with positive scalar permittivity $\varepsilon$, the most interesting case corresponds to positive quantities $\mu_\perp$ when wavenumbers

$$k_\perp = \frac{\omega}{c}\sqrt{\varepsilon\mu_\perp} \tag{48}$$

are real quantities. For given bias magnetic field $H_0$ and saturation magnetization $M_0$, $\mu_\perp > 0$ if $\omega < \omega_\perp$, where $\omega_\perp = \sqrt{\omega_0(\omega_0 + \omega_M)}$, $\omega_0 = \gamma H_0$, $\omega_M = \gamma 4\pi M_0$, and $\gamma$ is the gyromagnetic ratio.

For frequencies $\omega < \omega_\perp$ we consider now two separate regions: (a) $\omega < \omega_0$ and (b) $\omega_0 < \omega < \omega_\perp$. When $\omega < \omega_0$, one has $\mu_{eff_+} > 0$ and so there are real wavenumbers

$$k_{\|_\pm} = \frac{\omega}{c}\sqrt{\varepsilon\mu_{eff_\pm}} \,. \tag{49}$$

For $\omega_0 < \omega < \omega_\perp$, one has $\mu_{eff_+} < 0$. In this case

$$k_{\|_-} = \frac{\omega}{c}\sqrt{\varepsilon\mu_{eff_-}} \tag{50}$$

are real wavenumbers, while

$$k_{\|_+} = i\frac{\omega}{c}\sqrt{\varepsilon|\mu_{eff_+}|} \tag{51}$$

are imaginary wavenumbers.

A ferrite sample analyzed in Ref. [8] has the following material parameters: saturation magnetization $4\pi M_s = 1880$ G, the linewidth $\Delta H = 0.8$ Oe. permittivity $\varepsilon_r = 15$. For such material parameters, we will take a bias magnetic field $H_0 = 4900$ Oe and choose the cavity resonance frequency $f = 8.328$ GHz. This corresponds to the region where $\omega < \omega_0$. In neglect of lossless, a simple calculation gives the following quantities of wavenumbers in a ferrite medium: $k_\perp/k_0 = 6.2$, $k_{\|_+}/k_0 = 25.3$, and $k_{\|_-}/k_0 = 4.41$, where $k_0 = \omega/c$. A disk analyzed in Ref. [8] has diameter $D = 6$ mm and thickness $t = 0.5$ mm. For such geometry, it is evident that for the right-hand-circular-polarization waves propagating along a normal $z$ axis (with the wavenumber $k_{\|_+}$) there is almost the same order of the phase variation on a scale of the disk thickness as for the waves propagating along any direction in the $xy$ plane (with the wavenumber $k_\perp$) on a scale of a disk radius.

When a normally magnetized ferrite disk is placed in a rectangular waveguide cavity so that the disk axis is perpendicular to a wide wall of a waveguide (see Fig. 1), the phase velocities for plane waves propagating inside a disk are much less than outside a disk. An incoming wave propagates freely outside the ferrite, but can be trapped inside for some time. The mean "escape" time of a ray inside a ferrite disk into a waveguide vacuum space is much bigger than the time for the ray pass across the disk in a waveguide vacuum space. This clearly presumes a chaotic nature of the classical ray motion inside a sample with the time-reversal symmetry breaking effects. A ray is scattered several times from the disk's boundaries before exiting the disk. The electromagnetic fields approximately obey a statistical condition. This leads to special topological structures for the field and power flow distributions inside a ferrite disk. Fig. 2 shows a typical vortex picture for the Poynting vector distribution inside a ferrite disk when a disk is placed in a maximal cavity electric field. This picture was obtained based on numerical experiments with use of the HFSS (the software based on FEM method produced by ANSOFT Company) CAD simulation programs for 3D numerical modeling of Maxwell equations [27]. The program determines both modulus and phase of the fields. There is an evident picture of the Poynting-vector vortex in Fig. 2. A detailed study of the electromagnetic vortices inside a ferrite disk at frequencies $\omega < \omega_0$ is given in recent publications [7, 8].

Now let us consider the same disk with the same bias magnetic field $H_0 = 4900\,\text{Oe}$, but for the cavity frequency $f = 8.52$ GHz. This frequency is within the region $\omega_0 < \omega < \omega_\perp$. For a lossless ferrite, there are real quantities $k_\perp/k_0 = 6.33$ and $k_{\parallel-}/k_0 = 4.43$, but an imaginary quantity $k_{\parallel+}/k_0 = \text{j}34.5$. So in a direction of a bias magnetic field there are evanescent plane waves. Fig. 3 shows the Poynting vector distribution inside a disk for frequency $f = 8.52$ GHz, when a disk is placed in a maximal cavity electric field. There is a picture with a very chaotic Poynting-vector distribution, without any robust topological structure.

We will aim our further studies to an analysis of the role of the disk geometry. We consider a disk with the same diameter, but with a very small thickness, ten times less than in the previous case. So now the disk parameters are: diameter $D = 6\,mm$ and thickness $t = 0.05\,mm$. From numerical experiments with such a thin disk, we found that for the frequency region $\omega < \omega_0$ there are no qualitative differences with the pictures of the Poynting-vector and field distributions shown in Refs. [7, 8]. The situation, however, becomes drastically different for the frequency region $\omega_0 < \omega < \omega_\perp$. For a very thin disk in such a frequency region, one can observe now *topological resonant states*.

To a certain extent, electromagnetic wave processes inside a normally magnetized thin ferrite disk for a constant bias field and at frequencies $\omega_0 < \omega < \omega_\perp$ can be approximated as chaotically propagating plane waves in a vertically thin electromagnetic resonator with certain material filling. This allows reducing a problem to the quas-2D billiard model which mathematical properties are generally well studied for classical chaos and its quantum manifestation. For a 2D billiard system the random wave model implies that a typical chaotic wave function may be written locally as a random superposition of plane waves inside a ferrite at fixed frequency. The properties of membrane functions of individual ferrite-disk modes should be extracted from the behavior of whole families of ray trajectories in the *xy* plane. In the limit of small wavelength compared to the resonator size, one can use a well known Weyl formula for an average eigenvalue density [28]. On the other hand, the wave function in a disk may be expanded in circular waves with good angular moments and random amplitude coefficients [29].

## IV. CORRESPONDENCE BETWEEN ANALITICALLY DERIVED MDM VORTEX STATES AND NUMERICALLY MODELED ELECTROMAGNETIC VORTICES

Qualitatively, the field structures of topological resonance states in a thin ferrite disk at frequencies $\omega_0 < \omega < \omega_\perp$ are not strongly dependent on a disk diameter. So for our studies of the observed eigenfunction patterns we will use the disk having diameter $D = 3\,mm$ and thickness $t = 0.05\,mm$. Such disk geometry is very interesting for us since it corresponds to the geometry of a ferrite sample used in our recent experiments [20]. Based on the HFSS-program numerical studies, we analyze excitation of the topological resonant states in a ferrite disk placed in a microwave cavity at a bias magnetic field $H_0 = 4900\,Oe$. For our numerical studies we used a short-wall rectangular waveguide section. The disk axis was oriented along the *E*-field of a waveguide $TE_{10}$ mode (see Fig. 1). The ferrite material parameters are the same as for the above studies. Fig. 4 (a) shows numerically obtained frequency characteristic of an absorption coefficient for a ferrite disk in a waveguide cavity. One clearly sees the multiresonance regular absorption spectra. As we will show, for every absorption peak one has a robust picture of a topological resonant state. Moreover, we find a very good correspondence of these topological states with the analytically derived PFD vortices and the field patterns of MDMs in a ferrite disk.

The resonance peak positions obtained from an analytical solution of Eqs. (8) and (20) are shown in Fig. 4 (b). The numerical absorption peaks and the analytical peaks calculated based on the essential boundary conditions are, in fact, at the same positions [see Figs. 4 (a) and 4 (b)]. Figs. 5 and 6 show the power flow density distributions respectively for the 1$^{st}$ ($f$ = 8.52 GHz) and 2$^{nd}$ ($f$ = 8.66 GHz) modes in a quasi-2D ferrite disk. There is an excellent correspondence between a numerically modeled (HFSS-program) electromagnetic and analytically derived MDM power-flow-density vortices. Analytical results for the power flow density distributions were obtained based on Eq. (19) for the fundamental-thickness and first-order-azimuth MDMs. Numbers *n* in Eq. (19), being the numbers of zeros in the Bessel function, correspond to different radial variations. We analyzed distributions of $(p_n)_\theta$ for first two modes (*n* = 1, 2) at $\nu_n = +1$ when a bias magnetic field is directed along *z* axis.

A vortex can be defined as a circular flow which is attributed with a certain phase factor and a circular integral of a gradient of the phase gives a non-zero quantity. This quantity is multiple to a number of full rotations. In our case, the phase factor of the mode vortex appears from the topological properties of the azimuthally rotating magnetic field. Figs. 7 and 9 represent galleries of the numerically modeled magnetic field distributions on the upper plane of a ferrite disk, respectively, for the 1$^{st}$ and 2$^{nd}$ topological resonance states at different time phases. Every resonant state is characterized by a strong pronounced eigenfunction pattern. Similar distributions for the 1$^{st}$ and 2$^{nd}$ magnetic-dipolar modes analytically derived based on Eqs. (28) – (30) one can see, respectively, in Figs. 8 and 10. The observed pictures for magnetic fields of different modes have the same azimuth variations but are distinguished by the radius variations. A very peculiar property of these pictures is the fact of the *azimuthal rotation* of the mode magnetic field for an observer being situated outside a ferrite disk. Since for an observer being situated inside a ferrite disk, these modes have $2\pi$ azimuth variations, there is an evident $4\pi$ counterclockwise mode azimuthal rotation for an outside observer. The phase of the final mode state differs from that of the initial state by $\phi = \phi_d + \phi_g$, where $\phi_d$ and $\phi_g$ are the dynamical and geometrical phases, respectively [30].

At the same time, contrary to the magnetic field azimuth variations, there are no azimuth variations for the mode electric fields. This essentially differs from well known field distributions in integrable electromagnetic cylindrical structures. Really, in any integrable electromagnetic cylindrical structure (such as a cylindrical waveguide or resonator) one has from Maxwell's equations the same order of the azimuth variation for the electric and magnetic fields (see e.g. Refs. [31, 32]). Figs. 11 and 13 represent galleries of the numerically modeled electric field distributions on the upper plane of a ferrite disk, respectively, for the 1$^{st}$ and 2$^{nd}$ topological resonance states at different time phases. Figs. 12 and 14 show in-plane effective-electric-current distributions for the 1$^{st}$ and 2$^{nd}$ magnetic-dipolar modes analytically derived based on Eqs. (36) – (38). From these analytical results for effective-electric-current distributions one can obtain qualitative pictures for the electric fields. As it follows from Eq. (33), the electric field should be 90° shifted with respect to the effective electric current. This gives a good correspondence with the numerically modeled distributions for the electric fields.

It is very important to note that the in-plane electric fields on the upper and lower planes of a ferrite disk are in opposite directions at any time phase. Since the disk thickness is much, much less than the rectangular waveguide height, the disk in a cavity can be clearly replaced by a sheet with linear surface magnetic currents. A surface density of the effective magnetic current is expressed as

$$\vec{n} \times (\vec{E}_{upper} - \vec{E}_{lower}) = -\frac{4\pi}{c}\vec{i}^{\,m}, \qquad (52)$$

where $\vec{E}_{upper}$ and $\vec{E}_{lower}$ are, respectively, in-plane electric fields on the upper and lower planes of a ferrite disk and $\vec{n}$ is a normal to a disk plane directed along a bias magnetic field. Evidently, $\vec{E}_{upper} = -\vec{E}_{lower}$. Following pictures in Figs. 11 and 13, one can conclude that there are *rotating* linear surface magnetic currents.

**CONCLUSION**

In confined magnetically ordered structures one can observe vortices of magnetization and electromagnetic power flow vortices. There are topologically distinct and robust states. In this paper we showed that in a normally magnetized quasi-2D ferrite disk there exist eigen power-flow-density vortices of magnetic-dipolar-mode oscillations. Because of such circular power flows, the oscillating modes are characterized by stable magnetostatic energy states and discrete angular moments of the wave fields. Stable MDM vortex configurations are very attractive objects for fundamental physics studies of confined magnetic structures with dynamical symmetry breaking effects. The elements with such topologically distinct and robust states may find potential use as a unit cell for magnetic storage and magnetic logic devices. They can be also considered as very interesting objects for different microwave applications.

The observed MDM eigenstates and vortex structures in a thin ferrite disk can be analyzed in correspondence with numerically modeled Maxwell-equation objects in the ray phase space. Microwave systems with enclosed ferrite samples are considered as non-integrable Maxwellian problems. In a case of very thin ferrite disks one can observe topologically distinct and robust structures – the electromagnetic vortices – with eigenmode field structures. It is very interesting that in a quasi-2D ferrite disk one can find a clear correspondence between such pure numerical Maxwell-equation solutions of the vortex states and the analytically derived MDM vortex states obtained as a result of the MS-potential-wave-function spectral solutions.

**Figure captions**

Fig. 1. A model of a short-wall rectangular waveguide section with a normally magnetized ferrite disk.

Fig. 2. The Poynting vector distribution inside a ferrite disk at $\omega < \omega_0$. A bias magnetic field $H_0 = 4900$ Oe and the cavity resonance frequency $f = 8.328$ GHz. Disk diameter $D = 6$ mm and thickness $t = 0.5$ mm.

Fig. 3. The Poynting vector distribution inside a ferrite disk at $\omega_0 < \omega < \omega_\perp$. A bias magnetic field $H_0 = 4900$ Oe and the cavity resonance frequency $f = 8.52$ GHz. Disk diameter $D = 6$ mm and thickness $t = 0.5$ mm.

Fig. 4. Spectral characteristics for a thin ferrite disk. A bias magnetic field $H_0 = 4900$ Oe. Disk diameter $D = 3$ mm and thickness $t = 0.05$ mm. (*a*) Numerically obtained absorption coefficients; (*b*) analytically calculated peak positions.

Fig. 5. The power flow density distribution for the 1st mode (f = 8.52 GHz) in a quasi-2D ferrite disk. A bias magnetic field $H_0 = 4900$ Oe. Disk diameter $D = 3$ mm and thickness $t = 0.05$ mm. (a) Numerically modeled electromagnetic vortex; (b) analytically derived MDM vortex. A black arrow in Fig. 5 (a) clarifies the power-flow direction inside a disk.

Fig. 6. The power flow density distribution for the 2nd mode (*f* = 8.66 GHz) in a quasi-2D ferrite disk. A bias magnetic field $H_0 = 4900$ Oe. Disk diameter $D = 3$ mm and thickness $t = 0.05$ mm. (a) Numerically modeled electromagnetic vortex; (b) analytically derived MDM vortex. Black arrows in Fig. 6 (a) clarify the power-flow directions inside a disk.

Fig. 7. A perspective view for the numerically modeled magnetic field distributions on the upper plane of a ferrite disk for the for the 1st topological resonance state (*f* = 8.52 GHz) at different time phases. There are evident $4\pi$ azimuthal rotations. A bias magnetic field $H_0 = 4900$ Oe. Disk diameter $D = 3$ mm and thickness $t = 0.05$ mm.

Fig. 8. A gallery of the analytically derived in-plane magnetic field distributions on the upper plane of a ferrite disk for the 1st magnetic-dipolar mode (*f* = 8.52 GHz) at different time phases. A bias magnetic field $H_0 = 4900$ Oe. Disk diameter $D = 3$ mm and thickness $t = 0.05$ mm. There are evident $4\pi$ azimuthal rotations.

Fig. 9. A top view for the numerically modeled magnetic field distributions on the upper plane of a ferrite disk for the for the 2nd topological resonance state (*f* = 8.66 GHz) at different time phases. There are evident $4\pi$ azimuthal rotations. A bias magnetic field $H_0 = 4900$ Oe. Disk diameter $D = 3$ mm and thickness $t = 0.05$ mm.

Fig. 10. A gallery of the analytically derived in-plane magnetic field distributions on the upper plane of a ferrite disk for the 2nd magnetic-dipolar mode (*f* = 8.66 GHz) at some time phases. A bias magnetic field $H_0 = 4900$ Oe. Disk diameter $D = 3$ mm and thickness $t = 0.05$ mm. There are evident $4\pi$ azimuthal rotations.

Fig. 11. A top view for the numerically modeled electric field distributions on the upper plane of a ferrite disk for the for the 1st topological resonance state (*f* = 8.52 GHz) at different time phases. A bias magnetic field $H_0 = 4900$ Oe. Disk diameter $D = 3$ mm and thickness $t = 0.05$ mm.

Fig. 12. A gallery of the analytically derived in-plane effective-electric-current distributions on the upper plane of a ferrite disk for the 1st magnetic-dipolar mode (*f* = 8.52 GHz) at different time phases. A bias magnetic field $H_0 = 4900$ Oe. Disk diameter $D = 3$ mm and thickness $t = 0.05$ mm.

Fig. 13. A top view for the numerically modeled electric field distributions on the upper plane of a ferrite disk for the for the 2nd topological resonance state (*f* = 8.66 GHz) at different time phases. A bias magnetic field $H_0 = 4900$ Oe. Disk diameter $D = 3$ mm and thickness $t = 0.05$ mm.

Fig. 14. A gallery of the analytically derived in-plane effective-electric-current distributions on the upper plane of a ferrite disk for the 2nd magnetic-dipolar mode (*f* = 8.66 GHz) at some time phases. A bias magnetic field $H_0 = 4900$ Oe. Disk diameter $D = 3$ mm and thickness $t = 0.05$ mm.

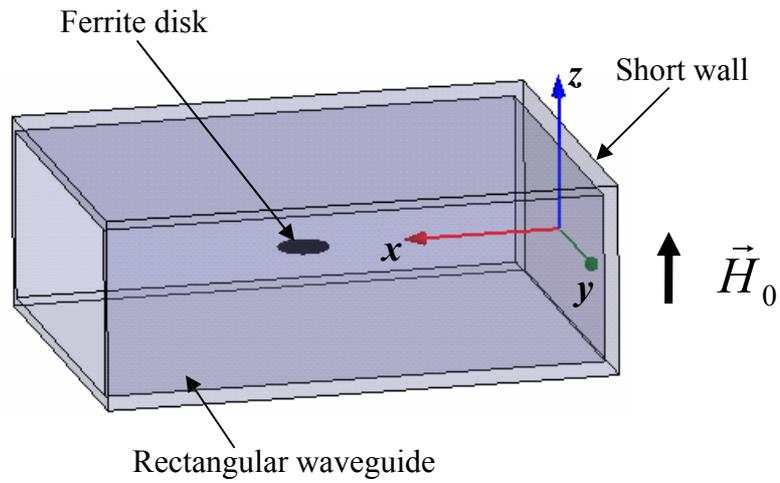

Fig. 1. A model of a short-wall rectangular waveguide section with a normally magnetized ferrite disk.

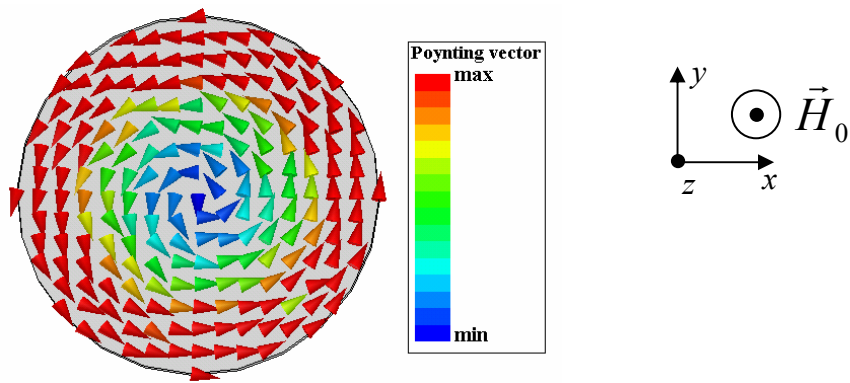

Fig. 2. The Poynting vector distribution inside a ferrite disk at $\omega < \omega_0$. A bias magnetic field $H_0 = 4900$ Oe and the cavity resonance frequency $f = 8.328$ GHz. Disk diameter $D = 6$ mm and thickness $t = 0.5$ mm.

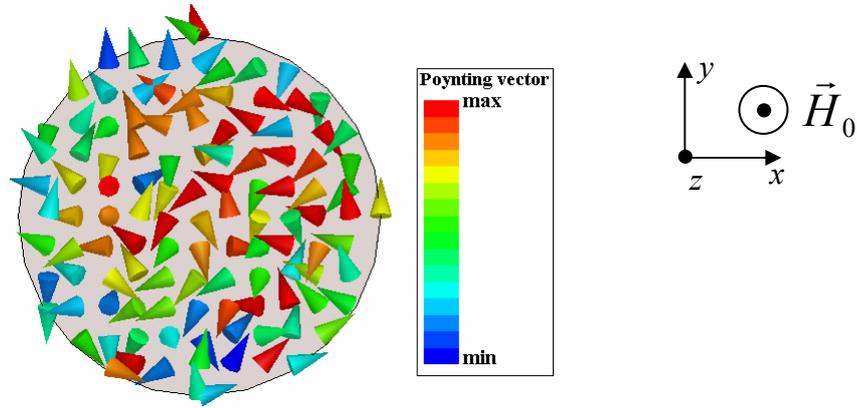

Fig. 3. The Poynting vector distribution inside a ferrite disk at $\omega_0 < \omega < \omega_\perp$. A bias magnetic field $H_0 = 4900$ Oe and the cavity resonance frequency $f = 8.52$ GHz. Disk diameter $D = 6$ mm and thickness $t = 0.5$ mm.

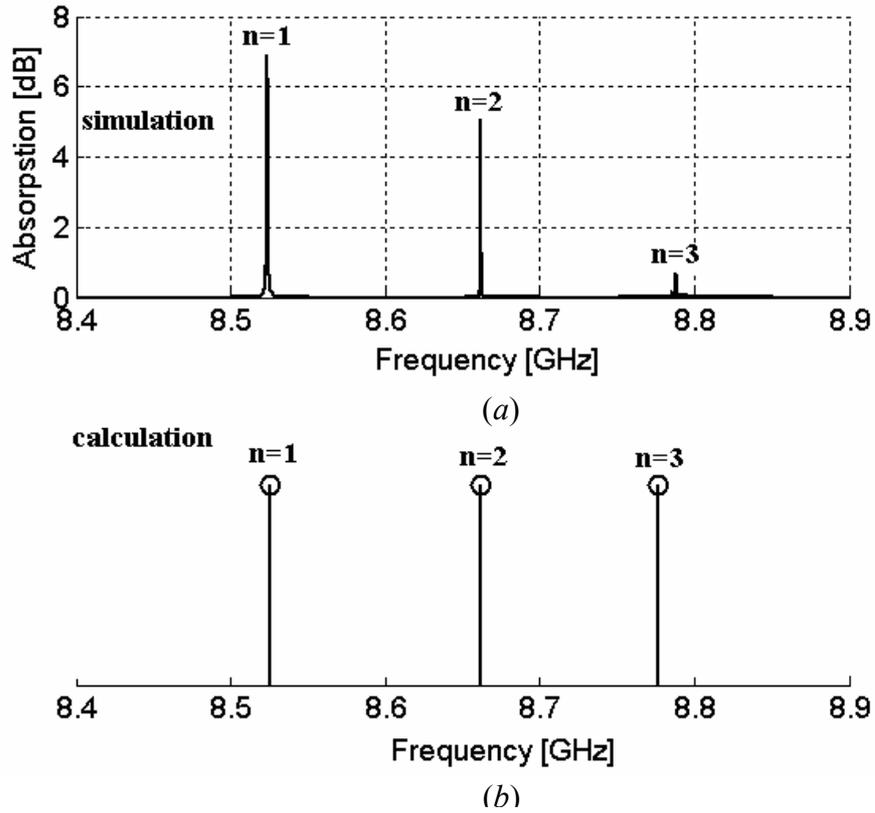

Fig. 4. Spectral characteristics for a thin ferrite disk. A bias magnetic field $H_0 = 4900$ Oe. Disk diameter $D = 3$ mm and thickness $t = 0.05$ mm. (*a*) Numerically obtained absorption coefficients; (*b*) analytically calculated peak positions.

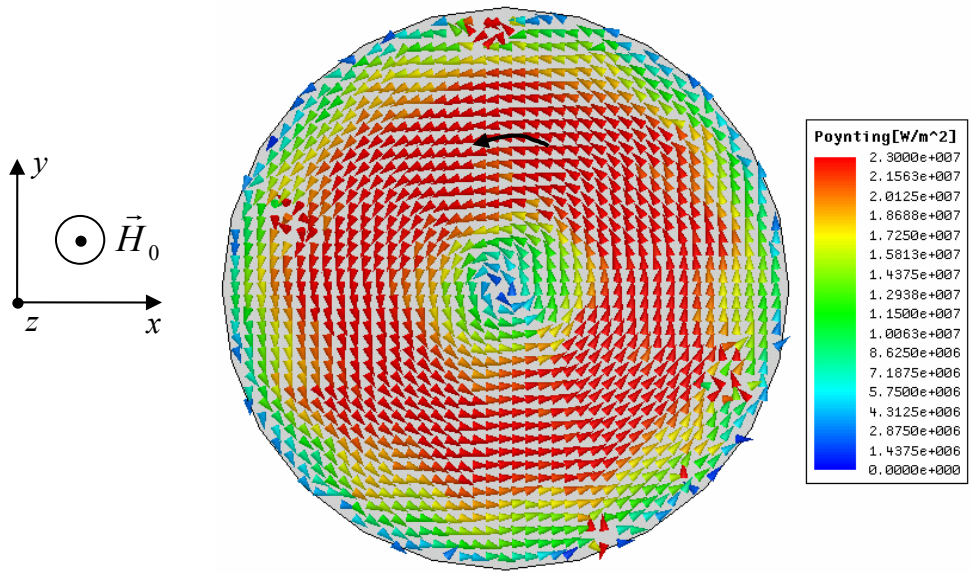

(a)

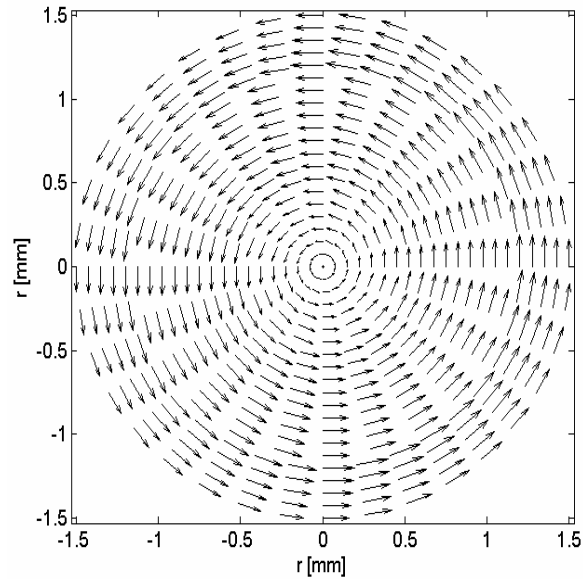

(b)

Fig. 5. The power flow density distribution for the 1$^{st}$ mode (f = 8.52 GHz) in a quasi-2D ferrite disk. A bias magnetic field $H_0 = 4900$ Oe. Disk diameter $D = 3$ mm and thickness $t = 0.05$ mm. (a) Numerically modeled electromagnetic vortex; (b) analytically derived MDM vortex. A black arrow in Fig. 5 (a) clarifies the power-flow direction inside a disk.

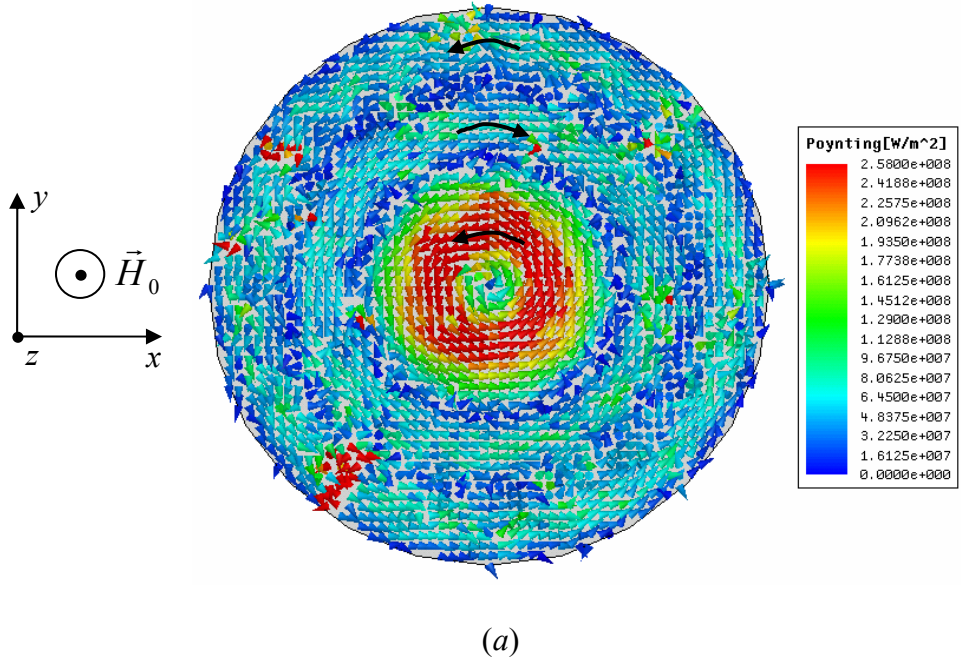

(*a*)

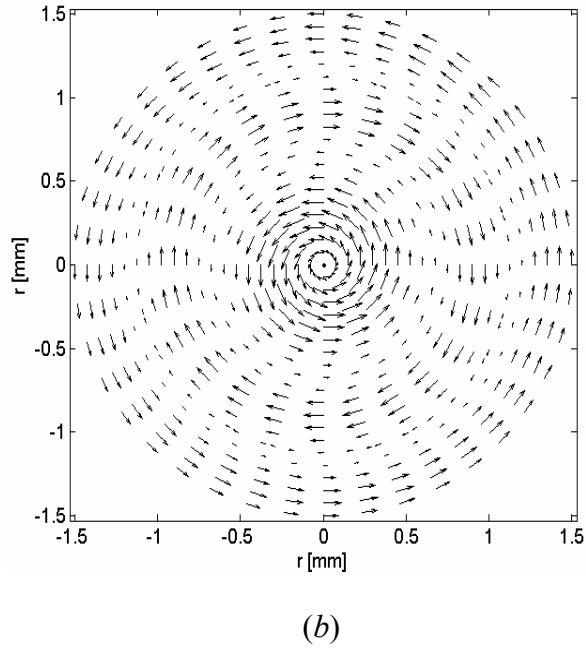

(*b*)

Fig. 6. The power flow density distribution for the 2$^{nd}$ mode ($f$ = 8.66 GHz) in a quasi-2D ferrite disk. A bias magnetic field $H_0 = 4900$ Oe. Disk diameter $D = 3$ mm and thickness $t = 0.05$ mm. (a) Numerically modeled electromagnetic vortex; (b) analytically derived MDM vortex. Black arrows in Fig. 6 (a) clarify the power-flow directions inside a disk.

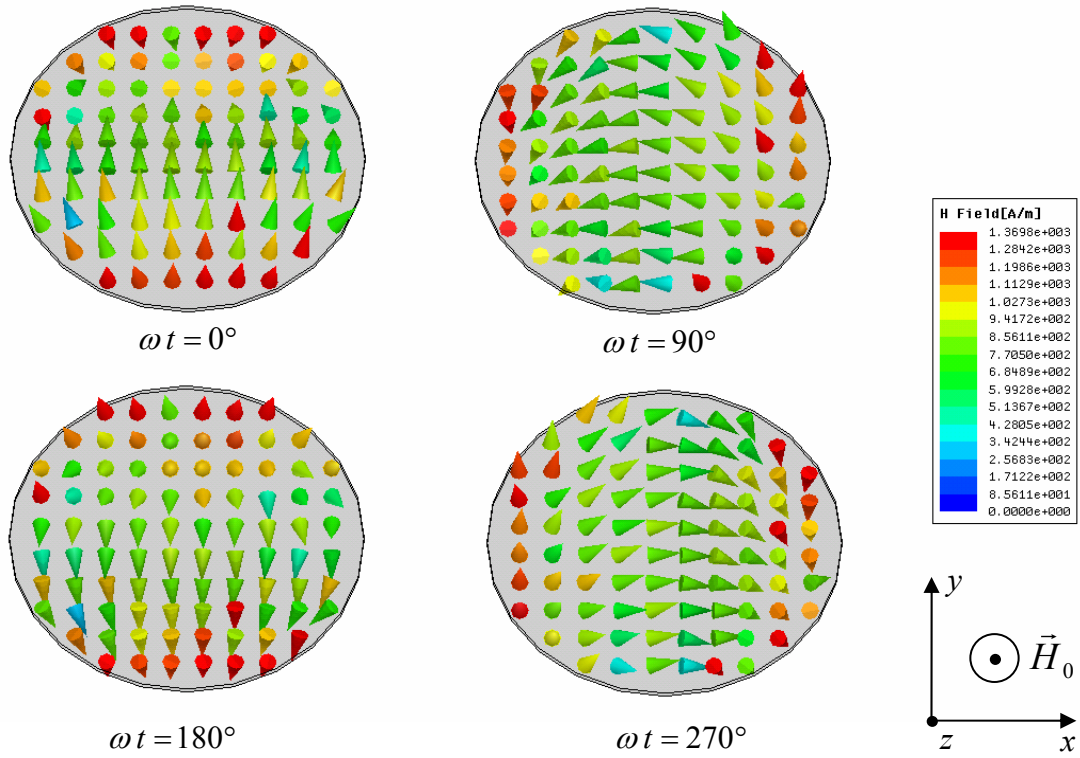

Fig. 7. A perspective view for the numerically modeled magnetic field distributions on the upper plane of a ferrite disk for the for the 1$^{st}$ topological resonance state ($f$ = 8.52 GHz) at different time phases. There are evident $4\pi$ azimuthal rotations. A bias magnetic field $H_0$ = 4900 Oe. Disk diameter $D$ = 3 mm and thickness $t$ = 0.05 mm.

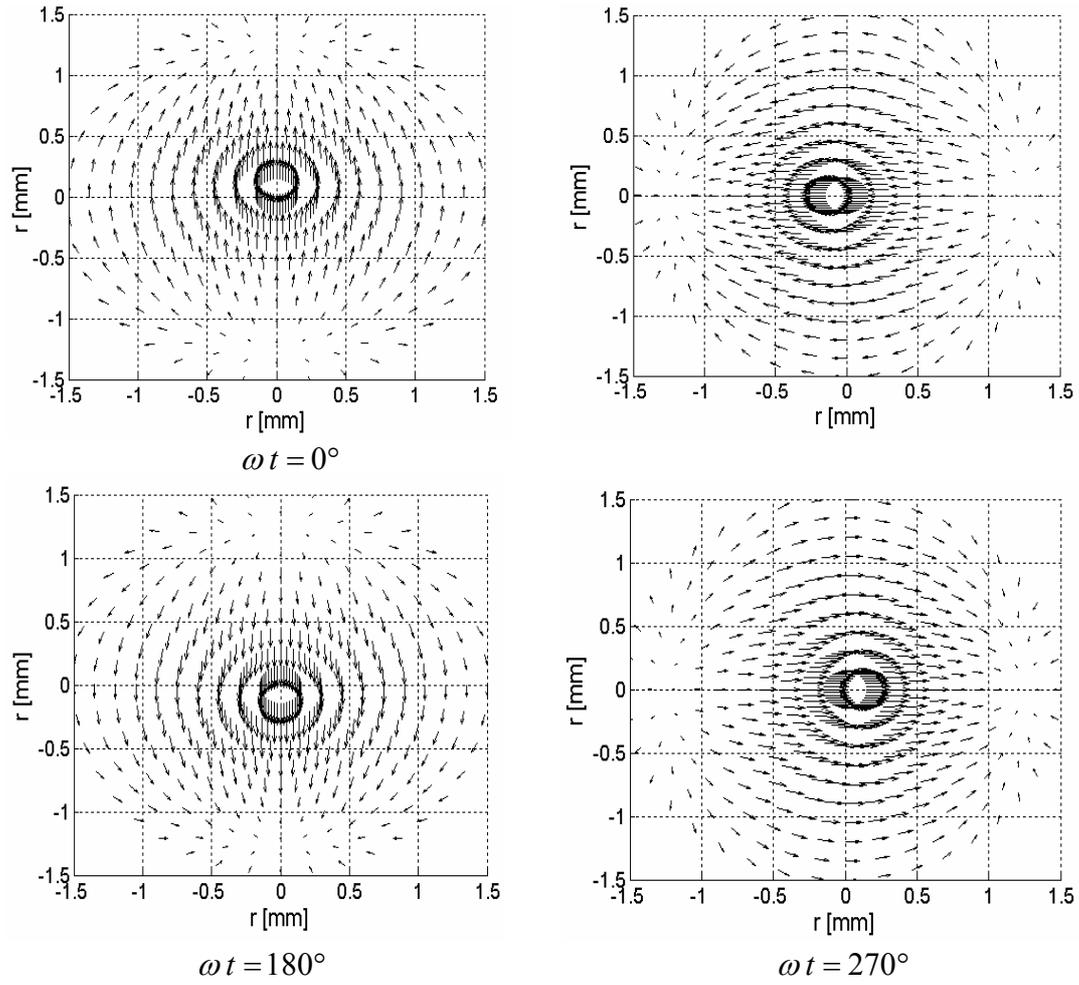

Fig. 8. A gallery of the analytically derived in-plane magnetic field distributions on the upper plane of a ferrite disk for the 1$^{st}$ magnetic-dipolar mode ($f$ = 8.52 GHz) at different time phases. A bias magnetic field $H_0$ = 4900 Oe. Disk diameter $D$ = 3 mm and thickness $t$ = 0.05 mm. There are evident $4\pi$ azimuthal rotations.

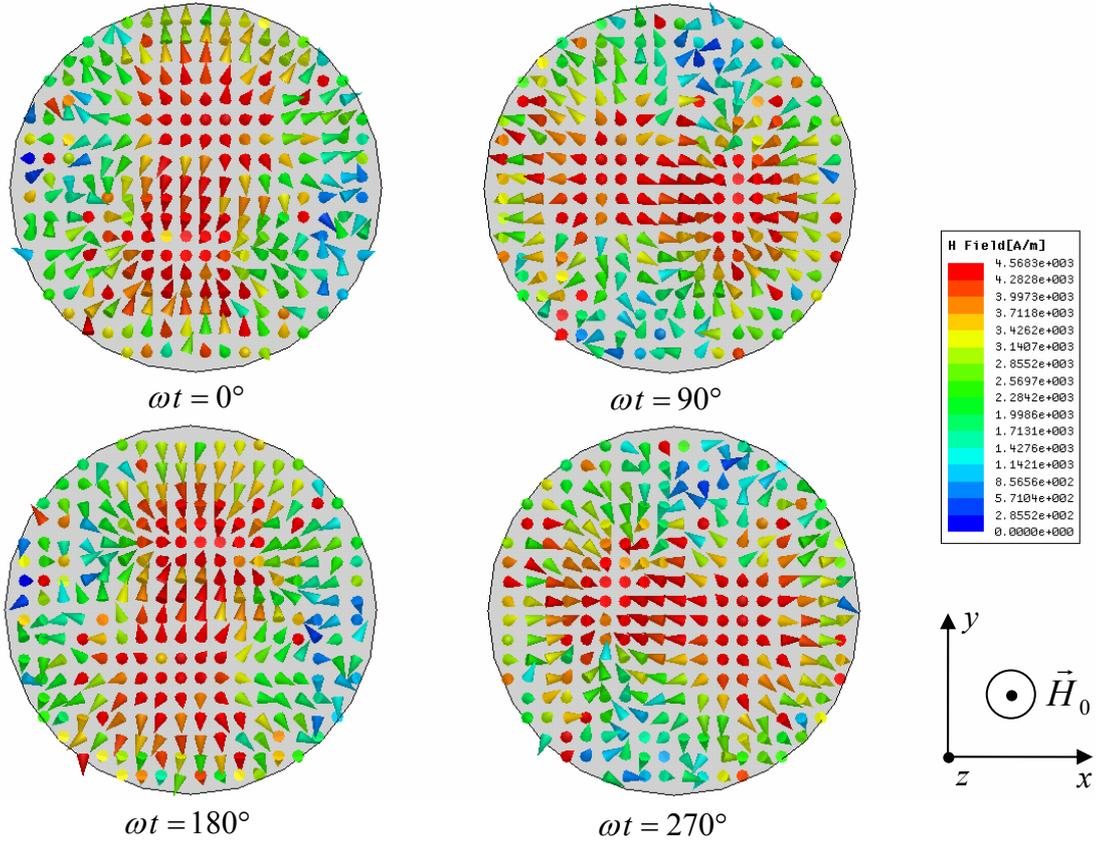

Fig. 9. A top view for the numerically modeled magnetic field distributions on the upper plane of a ferrite disk for the for the 2$^{nd}$ topological resonance state ($f$ = 8.66 GHz) at different time phases. There are evident $4\pi$ azimuthal rotations. A bias magnetic field $H_0 = 4900$ Oe. Disk diameter $D = 3$ mm and thickness $t = 0.05$ mm.

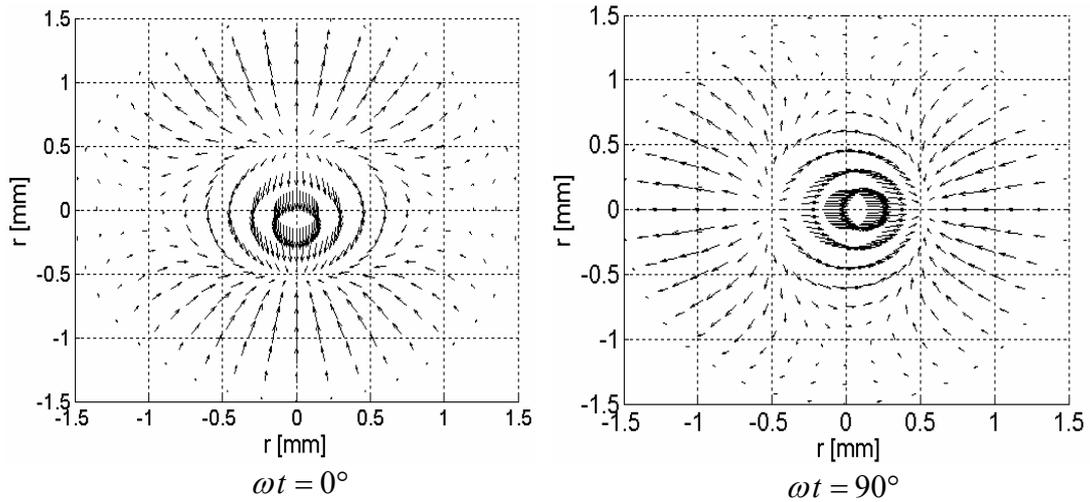

Fig. 10. A gallery of the analytically derived in-plane magnetic field distributions on the upper plane of a ferrite disk for the 2$^{nd}$ magnetic-dipolar mode ($f$ = 8.66 GHz) at some time phases. A bias magnetic field $H_0 = 4900$ Oe. Disk diameter $D = 3$ mm and thickness $t = 0.05$ mm. There are evident $4\pi$ azimuthal rotations.

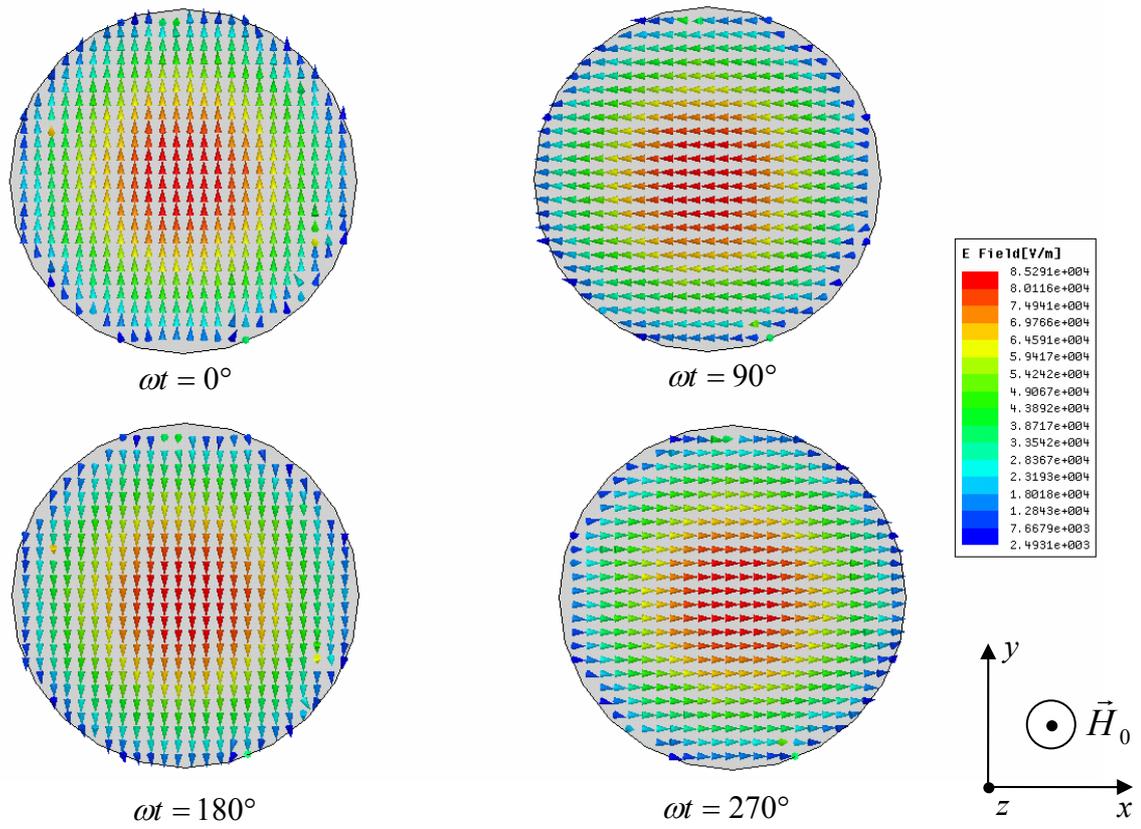

Fig. 11. A top view for the numerically modeled electric field distributions on the upper plane of a ferrite disk for the for the 1$^{st}$ topological resonance state ($f$ = 8.52 GHz) at different time phases. A bias magnetic field $H_0 = 4900$ Oe. Disk diameter $D = 3$ mm and thickness $t = 0.05$ mm.

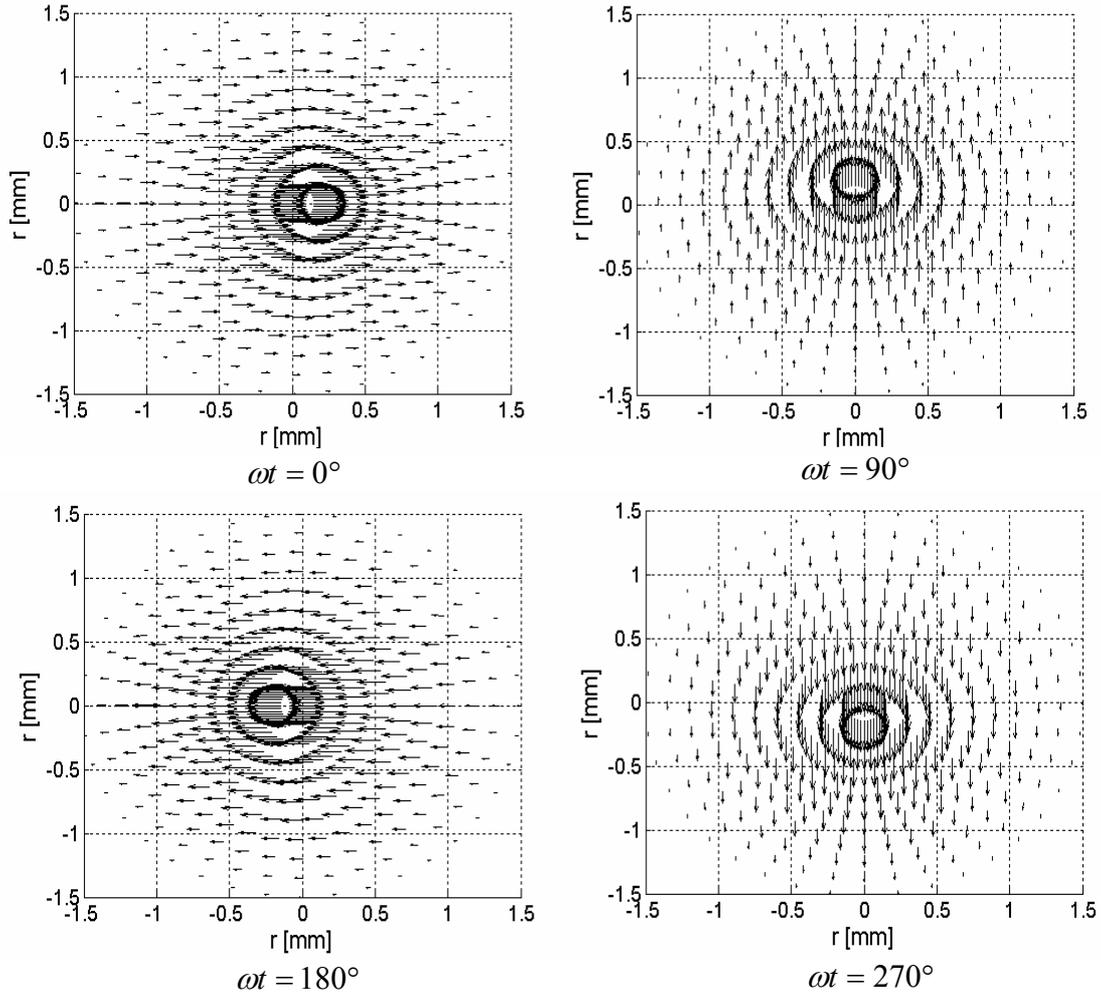

Fig. 12. A gallery of the analytically derived in-plane effective-electric-current distributions on the upper plane of a ferrite disk for the 1$^{st}$ magnetic-dipolar mode ($f$ = 8.52 GHz) at different time phases. A bias magnetic field $H_0$ = 4900 Oe. Disk diameter $D$ = 3 mm and thickness $t$ = 0.05 mm.

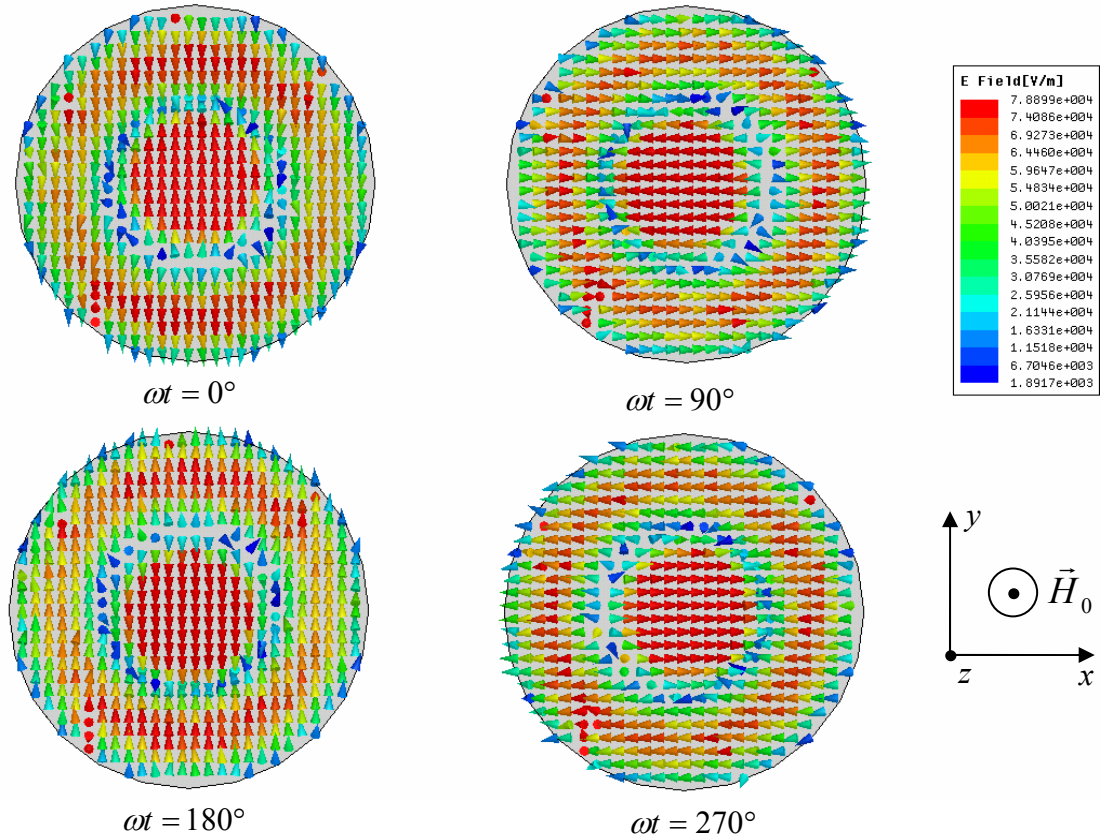

Fig. 13. A top view for the numerically modeled electric field distributions on the upper plane of a ferrite disk for the for the 2$^{nd}$ topological resonance state ($f$ = 8.66 GHz) at different time phases. A bias magnetic field $H_0 = 4900$ Oe. Disk diameter $D = 3$ mm and thickness $t = 0.05$ mm.

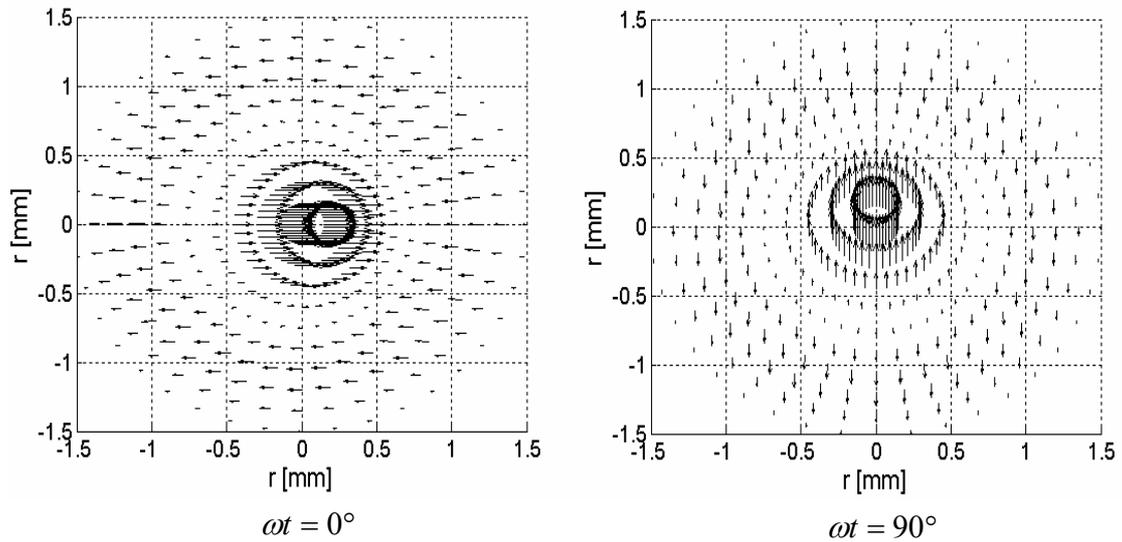

Fig. 14. A gallery of the analytically derived in-plane effective-electric-current distributions on the upper plane of a ferrite disk for the 2$^{nd}$ magnetic-dipolar mode ($f$ = 8.66 GHz) at some time phases. A bias magnetic field $H_0 = 4900$ Oe. Disk diameter $D = 3$ mm and thickness $t = 0.05$ mm.